\documentstyle[prl,aps,multicol]{revtex}
\begin{document}
\preprint{TNT 96-ShPre2Let-V.5}

\draft
\title{Growth of non-infinitesimal perturbations in turbulence}

\author{E. Aurell}
\address{Department of Mathematics, Stockholm University
          S--106 91 Stockholm, Sweden}

\author{G. Boffetta}
\address{Dipartimento di Fisica Generale, Universit\`a di Torino,
         Via Pietro Giuria 1, I-10125 Torino, Italy \\
         Istituto Nazionale Fisica della Materia, Unit\`a di Torino}

\author{A. Crisanti}
\address{Dipartimento di Fisica, Universit\`a di Roma ``La Sapienza'',
         P.le Aldo Moro 2, I-00185 Roma, Italy \\
         Istituto Nazionale Fisica della Materia, Unit\`a di Roma}

\author{G. Paladin}
\address{Dipartimento di Fisica, Universit\`a dell'Aquila,
         Via Vetoio, Coppito I-67100 L'Aquila, Italy \\
         Istituto Nazionale Fisica della Materia, Unit\`a dell'Aquila}

\author{A. Vulpiani}
\address{Dipartimento di Fisica, Universit\`a di Roma ``La Sapienza'',
         P.le Aldo Moro 2, I-00185 Roma, Italy \\
         Istituto Nazionale Fisica della Materia, Unit\`a di Roma}

\date{\today}

\maketitle

\begin{abstract}
We discuss the  effects of finite perturbations in fully developed turbulence
by introducing a measure of the chaoticity degree associated to a 
given scale of the velocity field.  This allows one to determine 
the predictability time for non-infinitesimal perturbations, 
generalizing the usual concept of maximum Lyapunov exponent. 
We also determine the scaling law for our indicator in the framework
of the multifractal  approach. 
We find that the scaling exponent is not sensitive to intermittency
corrections, but is an invariant of the multifractal models.
A numerical test of the results is performed in the shell model for 
the turbulent energy cascade.
\end{abstract}


\begin{multicols}{2}

\section*{}
The standard characterization of the chaotic behavior of a dynamical system 
is given by the maximum Lyapunov exponent $\lambda_{\max}$,
which measures the typical exponential rate of growth of an infinitesimal 
disturbance \cite{BeGaSt76}. It is thus expected that the 
predictability time is  proportional to $\lambda_{\max}^{-1}$, the shortest
characteristic time of the system.
The underlying point is that the growth of a perturbation 
is well described by the linear equations for the tangent vector
even if this cannot be literally true for non-infinitesimal perturbations.
There exist indeed many situations where the Lyapunov analysis has 
no relevance for the predictability problem and it is necessary to 
introduce indicators which are able to capture the essential features 
of a chaotic system. 
For instance, when two or more characteristic time scales are present
a direct identification of the Lyapunov and predictability times
leads to paradoxes as recently pointed out  in Ref. \cite{BofPaVul96}.

In this letter, we introduce a measure of the chaoticity 
degree related to the average doubling time that extends 
the concept of Lyapunov exponent in the case of 
non-infinitesimal perturbations.  
Our indicator is a scale-dependent Lyapunov exponent 
which becomes particularly useful when there exists a hierarchy 
of characteristic times such as the eddy turn-over times in three 
dimensional fully developed turbulence \cite{AuCrBofPaVu96}.

In turbulent flows it is natural to argue that the maximum Lyapunov 
exponent is roughly proportional the turnover time $\tau$ of eddies 
of the size of the Kolmogorov length $\eta$ (the viscous cut-off) 
that is the shortest characteristic time \cite{Ruel79}. 
By dimensional analysis, the turnover time of an eddy of size $\ell$ is
$\tau(\ell) \sim \ell^{1-h}$, where  $h$ is the scaling exponent of 
the velocity difference in the eddy 
\begin{equation}
v_{\ell} \equiv |\bbox{v}(\bbox{x}')-\bbox{v}(\bbox{x})| \sim \ell^h, \qquad
          \ell=|\bbox{x}'-\bbox{x}|.
\end{equation}
The viscous cut-off vanishes as a power of the Reynolds number 
${\rm Re}$, i.e. $\eta\sim {\rm Re}^{-1/(1+h)}$. These relations 
imply that the maximum Lyapunov exponent should scale as 
\begin{equation}
\label{eq:eq1}
\lambda_{\rm max} \sim 1\,/\,\tau(\eta) 
 \sim {\rm Re}^\alpha \quad {\rm with} \ \alpha={1-h\over 1+h}. 
\end{equation}

In the Kolmogorov K41 theory \cite{K41}, $h=1/3$ for all space points 
$\bbox{x}$ so that $\alpha=1/2$, as first pointed out by Ruelle \cite{Ruel79}.
However, the intermittency of energy dissipation leads to the existence 
of a spectrum of possible scaling exponents $h$ affecting the value of 
$\alpha$. In the multifractal approach \cite{Multifr}, the probability 
that the velocity difference on scale $\ell$ scales as 
$v_{\ell} \sim \ell^h$ is assumed to be $P_{\ell}(h)\sim \ell^{3-D(h)}$. 
This {\it ansatz} can be tested by measuring the scaling of the structure
functions
\begin{equation}
\langle v_{\ell} \rangle \sim \ell^{\zeta_P}.
\end{equation}
In the K41 theory \cite{K41,MoYa} $\zeta_p=p/3$ while in the multifractal 
scenario \cite{Multifr} $\zeta_p$ is a non-linear function of $p$
given by the Legendre transform of the function $D(h)$, 
$\zeta_p=\min_h \, [ \, h \, p - D(h) + 3 \, ]$.
Moreover, as a consequence of multifractality there is a spectrum
of viscous cut-offs \cite{PaVu87}, since each $h$ selects a different 
damping scale $\eta(h)\sim Re^{-1/(1+h)}$, and hence a spectrum of 
turnover times $\tau_{h}(\eta)$. 
To find the Lyapunov exponent, we have to integrate over the 
$h$-distribution \cite{CrJePaVu93}
\begin{eqnarray}
\lambda_{\rm max} &\sim& \int  \tau_h(\eta)^{-1} \, P_{\eta}(h)\, dh
 \nonumber \\
 &\sim& \int  \eta^{h-D(h)+2}  \,  dh \sim {\rm Re}^{\alpha}.
\label{eq:eq2}
\end{eqnarray}
In the limit ${\rm Re} \to \infty$, the integral can be estimated by 
the saddle point and gives $\alpha=\max_h \  [D(h)-2-h]/(1+h)$.
By using the function  $D(h)$ obtained with the random beta model  
fit \cite{Multifr,An84}  one has $\alpha=0.459..$.

In the predictability problem, we are interested in defining
the growth of an error on the velocity field. As usual we consider 
the Euclidean norm 
\begin{equation}
\label{eq:eq3}
\delta v(t)=\left(\int d^3x\, |\bbox{v}'(\bbox{x},t) - \bbox{v}(\bbox{x},t)|^2
            \right)^{1/2}.
\end{equation}
to introduce the notion of distance between two velocity fields
$\bbox{v}$ and  $\bbox{v}'$.

Then, the predictability time $T_{\rm p}$ is the time necessary
for an initial error  $\delta v(0)\equiv \delta_0$ to become larger 
than a given but  arbitrary threshold value $\Delta$:
\begin{equation}
\label{eq:eq4}
T_{\rm p}=\max_{t\ge 0} \, [ \ \delta v(t') \le \Delta \ {\rm for} \ t'<t].
\end{equation}
In a first approximation, neglecting the non-linear terms
of the evolution equation for the error growth and assuming that 
both $\delta_0$ and $\Delta$ are infinitesimal, one obtains
\begin{equation}
\label{eq:eq5}
T_{\rm p} \sim 
\lambda_{\rm max}^{-1} \, \ln (\Delta/|\delta_0|)
\approx \lambda_{\rm max}^{-1}.
\end{equation}
In turbulence, such a relation would imply that the predictability time 
decreases with the Reynolds number as ${\rm Re}^{-\alpha}$.
This is contradictory with the quite intuitive expectation that
the predictability time should be roughly proportional to the
turn-over time of the energy containing eddies on the large scales, 
and so practically independent of the Reynolds number \cite{Lor69,Leith72}.

The paradox stems from assuming the validity of the Lyapunov analysis
for perturbations $\delta v$ that are much larger than the typical 
velocity difference $v_{\eta} \sim \eta/\tau(\eta)$ 
on the Kolmogorov length scale $\eta$. In this case,
the error growth is non-exponential as it can be understood
by simple heuristic arguments \cite{Lor69,Leith72}.
The problem can be faced by generalizing the concept of 
maximum Lyapunov exponent to the case of non-infinitesimal perturbations.
The generalization is particularly useful 
in systems with many characteristic time-scales.

For this purpose, it is convenient to consider the time 
$T_{r}(\delta v)$ necessary for a perturbation to grow 
from $\delta v$ to $r \, \delta v$, for a generic $r>1$. 
For $r=2$ this is the doubling time  of a perturbation, usually studied  
in atmospheric predictability experiments \cite{Lor69,Leith72}.
After performing an average  over different realizations of the flow
or, equivalently, a time average along a trajectory $\bbox{v}(t)$
in the phase space, we introduce the scale-dependent Lyapunov exponent 
\begin{equation}
\label{eq:eq6}
\lambda(\delta v) = \left\langle{1\over T_{r}(\delta v)} \right\rangle
\ln r.
\end{equation}
Such a definition is consistent with the request of recovering the 
maximum Lyapunov exponent in the limit of infinitesimal error, since
\begin{equation}
\label{eq:eq7}
\lim_{\delta v \to 0} \lambda(\delta v)=\lambda_{\rm max}.
\end{equation}
It is easy to estimate the scaling of $\lambda(\delta v)$ when the 
perturbation is  in the inertial range 
$v_{\eta} \, \le  \delta v \le  \, v_{L}$, 
$L$ being the size of the energy containing eddies. 
In this case, following the phenomenological ideas of Lorenz, 
the doubling predictability time of an error of magnitude $\delta v$
can be identified with the turn-over time $\tau(\ell)$ of an eddy 
with typical velocity difference $v_{\ell} \sim \delta v$.  
Since $\tau(\ell) \sim \ell^{h-1} \sim v_{\ell}^{1-1/h}$, one has 
\begin{equation}
\label{eq:eq8}
\lambda(\delta v) \sim  \delta v^{-\beta}, \qquad
\beta=1/h-1.
\end{equation}
Neglecting intermittency, i.e. using the Kolmogorov value $h=1/3$, gives
$\beta=2$.
In the dissipative range $\delta v < v_{\eta}$, the error can be 
considered infinitesimal, implying $\lambda(\delta v) =\lambda_{\rm max}$. 

The intermittency of energy dissipation reflects the dynamical 
intermittency of the chaoticity degree, so that our arguments 
based on dimensional analysis cannot be fully correct.
In the framework of the multifractal approach, our indicator 
scales as 
\begin{equation}
\label{eq:eq9}
\lambda(\delta v) \sim \int dh \  \delta v^{[3 - D(h)]/h}\, \delta v^{1-1/h} 
\end{equation}
where we have used arguments similar to those leading to 
(\ref{eq:eq2}) and the scaling factor $\ell\sim \delta v^{1/h}$. 
>From the inequality $D(h)\le 3h + 2$, which is the analogous for 
turbulence of the standard inequality $f(\alpha)\le\alpha$ in 
multifractals, we have
\begin{equation}
\frac{2+h-D(h)}{h} \ge -2 \quad \mbox{for all}\ h.
\end{equation}
Equality holds for $h=h_{3}$, the exponent that realizes the minimum 
in the Legendre transform for the exponent of the third-order 
structure function 
$\zeta_3 = \min_{h}\left[ 3h + 3 - D(h) \right] = 1$.
Therefore a saddle point estimation of (\ref{eq:eq9}) gives
\begin{equation}
\label{eq:eq10}
-\beta =\min_{h} \left[\,\frac{2+h-D(h)}{h}\,\right] \, = -2.
\end{equation}

An important consequence of multifractality follows from 
the existence of a spectrum of dissipative cut-offs
$\eta(h)$ which reduces the effective inertial range 
where the scaling of $\lambda(\delta v)$ holds \cite{PaVu87,FrVe91}. 
To be more specific, the multifractal approach leads to 
$\eta(h)\sim {\rm Re}^{-1/(1+h)}$, and the integral
(\ref{eq:eq9}) has to be performed for 
$\widetilde{h}(\delta v)\leq h \leq h_{\rm max}$,
where $\widetilde{h}(\delta v)$ is given by
\begin{equation}
\delta v \sim {\rm Re}^{-\frac{\displaystyle \widetilde{h}(\delta v)}
                              {\displaystyle 1+\widetilde{h}(\delta v)}}.
\end{equation}
As a consequence, the scaling $\lambda(\delta v)\sim \delta v^{-\beta}$ 
holds only for $\delta v > {\rm Re}^{-h_{3} /(1+h_{3})}$, i.e., 
the inertial range is reduced by intermittency. In the range 
${\rm Re}^{-1/4} < \delta v < {\rm Re}^{-h_{3} /(1+h_{3})}$ we
expect a non-trivial shape of $\lambda(\delta v)$ depending on $D(h)$.

In order to test our results we have numerically studied the GOY
shell model \cite{GOY,JenPaVu91} for the energy cascade 
in fully developed turbulence. The model is an approximation
of the Navier-Stokes equations obtained by dividing the Fourier space into
shells of  wavenumbers $k_n<|\bbox{k}|< k_{n+1}$.
A complex scalar $u_n$ is associated with the $n^{\rm th}$
shell individuated by $k_n=k_0 \, 2^n$. It  represents
the velocity difference over a length scale $\ell\sim k_n^{-1}$.
Since the energy cascade in turbulence is believed to be 
local in the $k$-space with an exponentially 
decreasing interaction among shells, it is reasonable
to consider only the interactions of a shell with its nearest 
and next-nearest neighbors. The Navier-Stokes equations are then 
approximated by a set of ordinary differential equations:
\begin{eqnarray}
\label{eq:goy1}
 \frac{d}{dt}\,u_n &=& g_n -\nu k_n^2 u_n + f\delta_{n,4} \\
               g_n &=& i a_n k_n     u^*_{n+1} u^*_{n+2} +
                       i b_n k_{n-1} u^*_{n-1} u^*_{n+1} + \nonumber \\
                   & & i c_n k_{n-2} u^*_{n-2} u^*_{n-1}
\label{eq:goy2}
\end{eqnarray}
with $b_1= b_N= c_1= c_2= a_{N-1}= a_N= 0$. The coefficients of the 
nonlinear term obey $a_n + b_{n+1} + c_{n+2} = 0$ to ensure energy 
conservation for $f=\nu=0$. With the standard choice for three dimensional 
turbulence $a_n = 1$, $b_n = -1/2$ and $c_n = -1/2$, 
there is a second conserved quantity $\sum_n (-1)^n k_n |u_n|^2$, which 
in the shell model plays the role of helicity \cite{Kada}.

The shell model exhibits non-linear exponents $\zeta_p$
for the structure functions \cite{JenPaVu91}, as found in 
experimental data \cite{An84}. 
We have determined the scale dependent Lyapunov exponent by a numerical
integration of the GOY model starting form two different initial conditions. 
The distance between the two velocity fields is Euclidean, 
i.e. $\delta u=(\sum_n |u_n-u'_n|^2)^{1/2}$ and we have computed a time 
average over the trajectory $\{u_n(t)\}$ of the quantity 
$T_r(\delta u)^{-1}$ with $r=2^{1/2}$.

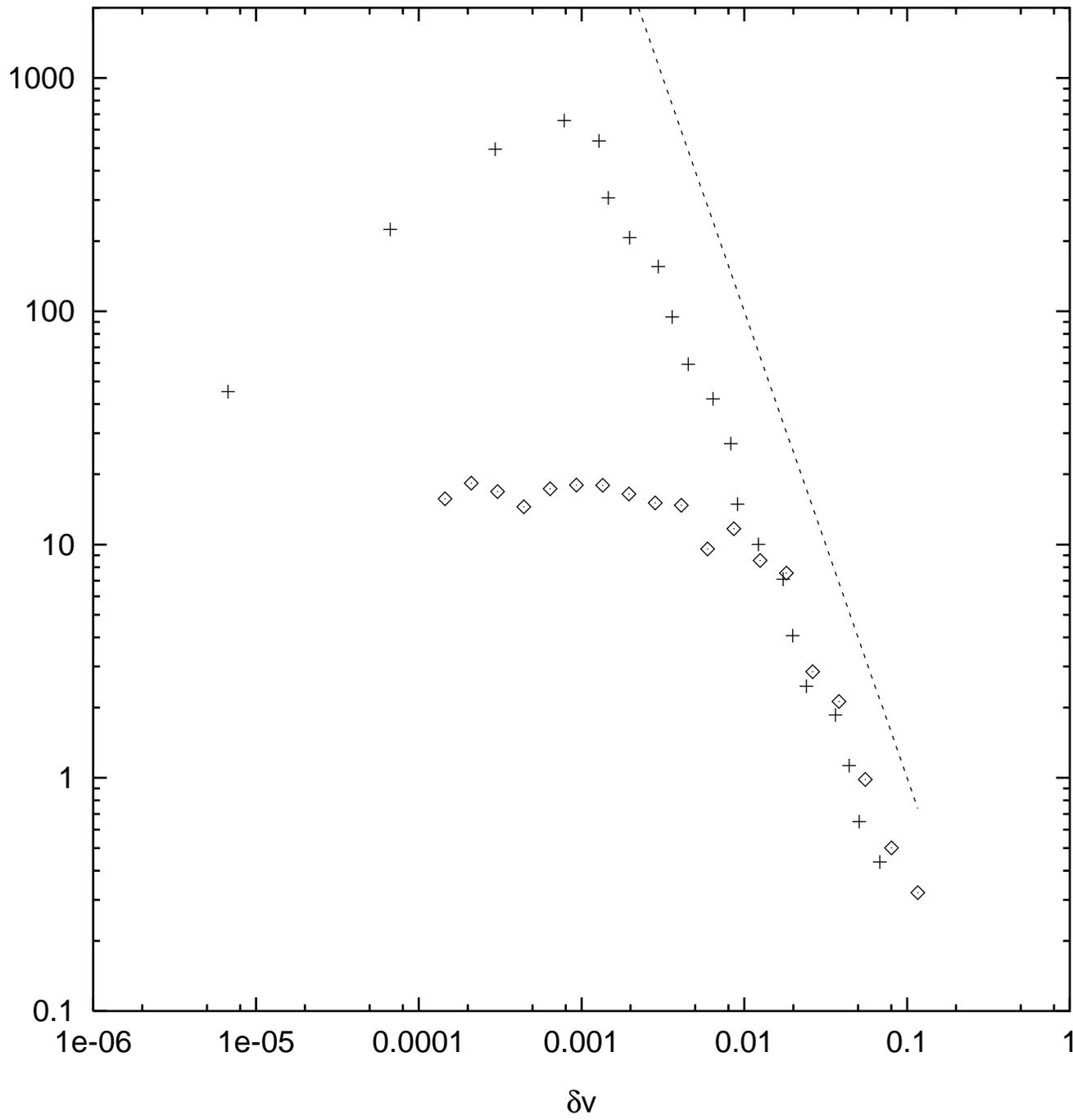
\begin{figure}
\caption{$\langle 1/T_r(\delta v)\rangle$ (diamond) as a function of $\delta v$
         for the GOY model
         with  $N=27$, $k_0= 0.05$, $f= (1+i)\times 0.005$ and $\nu=10^{-9}$.
         The crosses are the inverse of the eddy turn-over times
         $\tau^{-1}(\delta v)=k_n \,  \langle|u_n|^2\rangle^{1/2}$ 
         versus $\delta v=\langle|u_n|^2\rangle^{1/2}$. 
	 The straight line has slope $-2$.
}
\label{fig:f1}
\end{figure}

Figure \ref{fig:f1} shows the scaling of $\langle 1/T_r(\delta v)\rangle$ 
as a function of $\delta v$ in the GOY model with $N=27$ shells and 
viscosity $\nu=10^{-9}$. For comparison we have also computed the eddy 
turn-over times
\begin{equation}
\tau_n^{-1}= k_n \langle|u_n|^2\rangle^{1/2}.
\end{equation}
We thus provide a clear evidence that there is a large range of small scales  
where $\lambda(\delta u)=\lambda_{\rm max}$ and $\tau_n \sim u_n^{-\beta}$, 
as a consequence of the reduction of the inertial range for the scale
dependent Lyapunov exponent. 
Note that in the GOY model 
$\lambda_{\rm max} \sim 10^{-2} \, \tau(\eta)^{-1}$, though the
dependence of the two quantities on the Reynolds number is the same. 

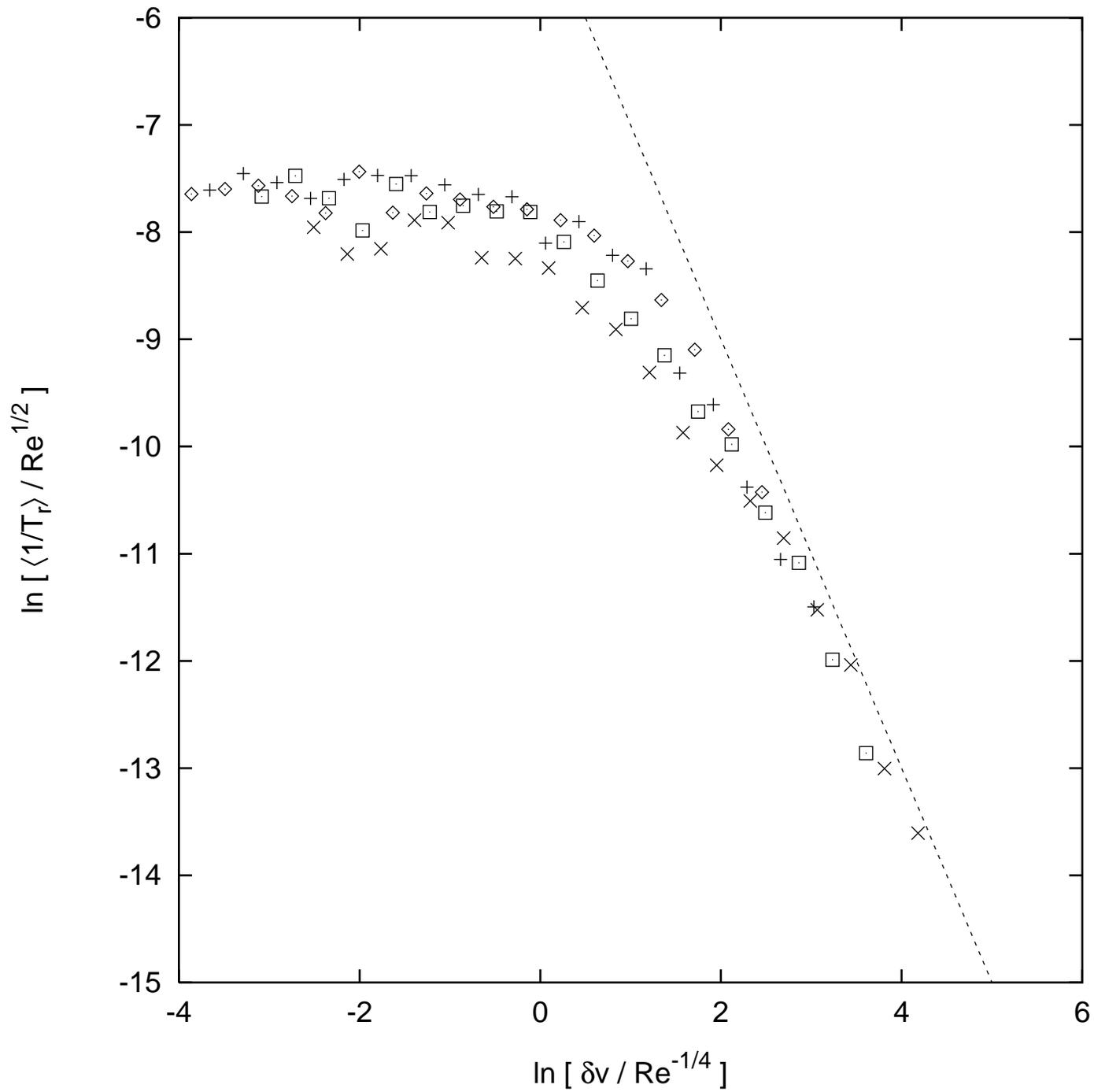
\begin{figure}
\caption{$\ln\left[\langle 1/T_r(\delta v)\rangle/{\rm Re}^{1/2}\right]$ 
	 versus $\ln\left[\delta v/{\rm Re}^{-1/4}\right]$
         at different Reynolds numbers ${\rm Re}=\nu^{-1}$. 
         The results are obtained in the GOY model for
         $k_0= 0.05$, $f= (1+i)\times 0.005$ and:
	 (diamond) $N = 24$ and $\nu=10^{-8}$;
	 (plus) $N = 27$ and $\nu=10^{-9}$;
	 (square) $N = 32$ and $\nu=10^{-10}$;
	 (cross) $N = 35$ and $\nu=10^{-11}$.
	 The straight line has slope $-2$.
}
\label{fig:f2}
\end{figure}

In Figure \ref{fig:f2}, we have plotted the rescaled quantity
$\lambda(\delta u)/Re^{\alpha}$ versus $\delta u/Re^{-\gamma}$
using the Kolmogorov values $h=1/3$, $\alpha=(1-h)/(1+h)=1/2$ 
and $\gamma=h/(1+h)=1/4$. The collapse of the data obtained at 
different Reynolds numbers is a pretty confirmation of our scaling arguments. 
The reduction of the inertial range for the scale dependent Lyapunov 
exponent reveals itself in the small range where 
$\lambda(\delta v)\sim \delta v^{-2}$ holds.

We conclude by noting that our scale-dependent Lyapunov exponent 
$\lambda(\delta v)$ has some similarity with the concept of the 
$\epsilon$-entropy recently discussed by Gaspard and Wang 
\cite{ShaWea49,Kol56,GaspWa} for the treatment of experimental data. 
We stress that in chaotic systems the maximum Lyapunov exponent 
is often more relevant and more easily computed than the 
Kolmogorov-Sinai entropy. Therefore we believe that our $\lambda(\delta v)$
will often be more relevant and more easily computable than the 
$\epsilon$-entropy. Moreover, since we use the evolution law and 
not experimental data, in our case there are no particular limitations 
as to the number of degrees of freedom involved.

In conclusion, when the perturbations are non-infinitesimal 
it is necessary to extend the definition of Lyapunov exponent to make it 
physically consistent. The generalization proposed in this letter
is particular useful when many characteristic time scales are present.
Our result allows one to get a quantitative control of the growth of
perturbations which are non-infinitesimal, looking at the average of 
the inverse doubling time. By this definition one has the two advantages 
of maintaining the link with the forecast limitation of a system  and of 
recovering  the maximum Lyapunov exponent in the limit of infinitesimal 
perturbations. The scale-dependent Lyapunov exponent thus is an important 
tool of investigation of highly dimensional dynamical systems 
and, far from being limited to the predictability problem
of turbulent flows in geophysics \cite{Lor69,Leith72,Lor82}, it 
can assume a great relevance in the characterization 
of very different chaotic phenomena.

This work was supported by the Swedish Natural Science
Research Council through grant S-FO-1778-302 (E.A.),
and by INFN {\it Iniziativa Specifica FI11} (G.B., G.P.  and A. V.).
E.A. thanks Dipartimento di Fisica, Universit\`a di Roma
``La Sapienza'' for hospitality.
G. B. thanks the ``Istituto di Cosmogeofisica del CNR'', Torino, for
hospitality.


\narrowtext

\end{multicols}

\begin{references}

\bibitem{BeGaSt76}
        G. Benettin, L. Galgani and J.M. Strelcyn,
        Phys. Rev. A {\bf 14}, 233 (1976).

\bibitem{BofPaVul96}
        G. Boffetta, G. Paladin and A. Vulpiani, 
        J. Phys. A, (1996) in press.

\bibitem{AuCrBofPaVu96}
        E. Aurell, A. Crisanti, G. Boffetta, G. Paladin and
        A. Vulpiani,
	Phys. Rev. E (1996) in press.

\bibitem{Ruel79}
        D. Ruelle,
        Phys. Lett. {\bf 72A}, 81 (1979).

\bibitem{K41} 
        A.N. Kolmogorov,
        C.R. (Dokl.) Acad. Sci. USSR {\bf 30}, 301 (1941).

\bibitem{Multifr}
        R. Benzi, G. Paladin, G. Parisi and A. Vulpiani, 
        J. Phys. A {\bf 17}, 3521 (1984);
        G. Parisi and U. Frisch , 
        in  {\it Turbulence and Predictability of Geophysical
           Flows and Climatic Dynamics}  edited M. Ghil et al.
           (North-Holland, New York, 1985), pag. 84;
        G. Paladin and A. Vulpiani, 
        Phys. Rep. {\bf 156}, 147 (1987).

\bibitem{MoYa} 
        A.S. Monin and A.M. Yaglom, 
        {\it Statistical Fluid Mechanics}
        (MIT, Cambridge, MA, 1975),  Vol II.

\bibitem{PaVu87}
        G. Paladin and A. Vulpiani, 
        Phys. Rev. {\bf A35}, 1971 (1987).

\bibitem{CrJePaVu93}
        A. Crisanti, M.H. Jensen, A. Paladin and A. Vulpiani,
        Phys. Rev. Lett. {\bf 70}, 166 (1993);
        A. Crisanti, M.H. Jensen, A. Paladin and A. Vulpiani,
        J. Phys. A {\bf 26}, 6943 (1993).

\bibitem{An84}
        F. Anselmet, Y. Gagne, E. J. Hopfinger and R. Antonia,
        J. Fluid Mech. {\bf 140}, 63 (1984).

\bibitem{Lor69}
        E.N. Lorenz Tellus {\bf 21}, 3 (1969)

\bibitem{Leith72}
        C.E. Leith and R. H. Kraichnan,
        J. Atmos. Sci. {\bf 29}, 1041 (1972);
        D.K. Lilly, 
        in  {\it Dynamic Meteorology} pag. 353 Ed. P. Morel
        (D. Reidel Publishing Company, Boston 1973).

\bibitem{FrVe91}
        U. Frisch and M. Vergassola,
        Europhys. Lett. {\bf 14}, 439 (1991).

\bibitem{GOY}
        E.B. Gledzer, Sov. Phys. Dokl. {\bf 18},  216 (1973);
        M. Yamada and K. Ohkitani,
        J. Phys. Soc. Jap. {\bf 56}, 4210 (1987);
        M. Yamada and K. Ohkitani,
        Prog. Theor. Phys. {\bf 79}, 1265 (1988).

\bibitem{JenPaVu91}
        M.H. Jensen, G. Paladin and A. Vulpiani, 
        Phys. Rev. A {\bf 43}, 798 (1991).

\bibitem{Kada}
        L.P. Kadanoff, D. Lohse, J. Wang and R. Benzi,
        Phys. Fluid A {\bf 7}, 617 (1995).

\bibitem{ShaWea49}
	C. Shannon and W. Weaver, {\it The mathematical theory
	of communication}, The University of Illinois Press, Urbana
	(1949).

\bibitem{Kol56}
	A.N. Kolmogorov, IRE Trans. Inf. Theory {\bf 1}, 102 (1956).

\bibitem{GaspWa}
        P. Gaspard and X.-J. Wang, Phys. Rep {\bf 235}, 291 (1993);
        X.-J. Wang and P. Gaspard , Phys. Rev. A {\bf 46}, R3000 (1992).

\bibitem{Lor82}
        E.N. Lorenz Tellus {\bf 34}, 505 (1982).
\end{references}
\end{document}